\documentclass[preprint,showpacs,preprintnumbers,amsmath,amssymb]{revtex4}
\usepackage{graphicx}
\usepackage{dcolumn}
\usepackage{bm}
\usepackage{amssymb}
\usepackage{amsmath}
\usepackage{latexsym}

\begin{document}

\title{On the evolution of compact binary black holes}

\author{G.G. Adamian$^{1}$,   N.V. Antonenko$^{1}$,   H. Lenske$^{2}$, and V.V. Sargsyan$^{1}$}

\affiliation{
$^{1}$Joint Institute for Nuclear Research, 141980 Dubna, Russia,\\
$^{2}$Institut f\"ur Theoretische Physik der
Justus--Liebig--Universit\"at, D--35392 Giessen, Germany}

\date{\today}

\begin{abstract}
Based on the consideration of potential energy of the di-black-hole as a function of mass asymmetry (transfer) collective coordinate,
the possibility of matter transfer between the black holes in a binary system is investigated.
The sensitivity of the calculated results is studied to the value  of the total mass of binary system.
The conditions for the merger of two black holes are analyzed in the context of gravitational wave emission.
\end{abstract}


\maketitle

\section{Introduction}\label{sec:Intro}
The general point of view is that
the binaries of compact object, composed of white dwarfs, neutron stars and
black holes, eventually merge through gravitational wave emission,
as observed recently with the LIGO and
Virgo interferometers \cite{Abbott}. There is also a general consensus that the
physical processes  during the merger belong to the biggest unsolved problems of stellar evolution.

Mass transfer is an important process for close
binary systems, because of its decisive role for the fate of the system. Hence, it is meaningful and necessary to study the  evolution of binary stellar
systems in the mass asymmetry coordinate
$\eta = (M_1-M_2)/M$ [$M_k$ ($k$=1,2) are the object
masses and $M=M_1+M_2$ is the total mass of system].
In our previous works \cite{IJMPE,IJMPE2,IJMPE4}, we have analyzed the
total potential energy $U(\eta)$ and the orbital period as
functions of $\eta$ at fixed total mass $M=M_1+M_2$
and orbital angular momentum $L$ of the contact binary stars and galaxies.
We have shown that the mass asymmetry (transfer) collective degree of freedom plays
a comparable important role   in macroscopic objects, being of comparable importance
as in microscopic dinuclear systems. In close binary galaxies and stars,
the   mass asymmetry coordinate
governs the asymmetrization (the transfer of mass from the lighter component to the heavy one)
and symmetrization (the transfer of mass from the heavier component to the light one)
of the system. As found, the symmetrization of binary object is the most favorable
evolution channel.
The symmetrization of binary galaxies and stars leads to the release of a
large amount of energy. For example, in the case of binary galaxies, the released energy is about $10^{48-52}$~J,
thus reaching the energy release of novae or even close to supernovae events \cite{IJMPE2}.
Thus, the symmetrization of close binary galaxy  due to the mass transfer is one of the important
sources of the transformation of the gravitational  energy  to other types of energy,
like radiation energy, in the Universe.

By this article, we intend to add a few aspects highlighting the role of mass transfer for binary black holes (BBH). Under all stellar binaries, BBH are undoubtfully a special class of objects. The presence of the event horizon inhibits the flow of mass between the core systems. On a speculative level, quantum tunneling as the Hawking radiation \cite{Hawking} may play a faint, but most likely negligible, role in view of the yet pending firm confirmation. However, BBH are embedded typically in a cloud of remnant matter, left over from the progenitor stars. Hence, a BBH system must always be considered together with the surrounding accretion disk. The material of that disk may serve for mass transfer between the core systems by the so--called \emph{sloshing effect} as illustrated clearly by the hydrodynamical calculations in Ref. \cite{Bowen:2016mci}.

In order to understand the conditions for matter exchange in a BBH system, we investigate the potential energy of a di-black-hole system as a function of mass asymmetry coordinate $\eta$.
We consider
the possibility of this transfer based on the calculation of the  potential energy of BBH system -- always to be understood together with its accretion disk -- is calculated as a function of mass asymmetry. Note
that in addition to its own extreme strong gravitational fields of two black holes, the potential energy contains the interaction energy between two black holes. We use Newtonian mechanics, being well aware of the strong gravitational fields. As found in \cite{Bowen:2016mci}, at the distance post--Newtonian effects may alter the binary potential not more than 25\% which will not affect the overall BBH properties. Of course, general relativity will become essential at separations close to touching which is a scenario beyond the scope of this work.

In section \ref{sec:Theory} we discuss the theoretical approach and the derived predictions, based on a fix point analysis of the dependence of the total BBH energy functional on the mass asymmetry coordinate. The work is summarized in section \ref{sec:Summary}.

\section{Theoretical Method and Its Application}\label{sec:Theory}
%

The total potential energy of the BBH system
\begin{eqnarray}
U=U_1+U_2+V_{1R}+V_{2R}+V
\label{eq_pot}
\end{eqnarray}
is given by the sum of the potential  $U_k$ ($k=1,2$) and  rotational energies $V_{kR}$ ($k=1,2$)
of the two nonzero spin black holes, and black-hole-black-hole interaction potential $V$.
The energy of the black hole  "$k$" is
\begin{eqnarray}
U_{k}=-\frac{G M_k^2}{R_k},
\label{eq_pot22}
\end{eqnarray}
where $G$, $M_k$, and $R_k$ are the gravitational constant, mass, and radius of the black hole, respectively.
%
The rotation energy of black-hole is
\begin{eqnarray}
V_{kR}=\frac{S_k\omega_k}{2}=\frac{M_{k}c^2}{2},
\label{eq_pot1R}
\end{eqnarray}
where
$S_{k}=M_{k}R_{k}c$  and
$\omega_k=c/R_k$
are the spin
and   rotational frequency of the  black hole, respectively. Here, $c$ is the speed of light.

The  radius  of the event horizon (distance from the gravitating mass $M_k$ at which the particle velocity
becomes equal to the speed of light  $c$) \cite{Murad1,Murad2,Murad3}
\begin{eqnarray}
R_k=\frac{GM_k}{c^2}
\label{eq_pot22kerr}
\end{eqnarray}
 is derived from the energy conservation law for a particle with mass $m$: $-GM_km/R_k+mc^2=0$,
where $mc^2$ is the sum of kinetic $mc^2/2$ and rotation $mc^2/2$ energies.
Note that the derivation of the radius $R_k$ is  based on classical mechanics and Newtonian law of gravity.
Similarly, in Ref. \cite{Michell}, an expression for the radius  $R_k=2GM_k/c^2$ was obtained  in the case of a non-rotating black-hole.
Employing  Eqs. (\ref{eq_pot22})  and   (\ref{eq_pot22kerr}),
we obtain
\begin{eqnarray}
U_{k}&=&- M_{k}c^2.
\label{eq_pot1}
\end{eqnarray}

Because the two black holes rotate with respect to each other around the common center of mass,
the black-hole-black-hole interaction potential $V(R)$ contains, together with the gravitational energy
of  interaction $V_G$ of two black holes,  the kinetic energy of orbital rotation $V_R$:
\begin{eqnarray}
V(R)=V_G+V_R=V_G+\frac{L^2}{2\mu R^2}=V_G+\frac{\mu v^2}{2},
\label{eq_pot3}
\end{eqnarray}
where  $v=(GM[2/R-1/R_m])^{1/2}$ and $R_m$ are the speed
and   the semimajor axis of  the elliptical relative orbit,
respectively,
$L$ is the orbital angular momentum of the  di-black-hole system,  and
$\mu=M_1M_2/(M_1+M_2)=M_1M_2/M$ is the reduced mass.
At $R\ge R_t=R_1+R_2$, the black-hole-black-hole interaction potential is
\begin{eqnarray}
V_G(R)=-\frac{GM_1M_2}{R}.
\label{eq_pot3n}
\end{eqnarray}
From the conditions $\partial V/\partial R|_{R=R_m}=0$ and  $\partial^2 V/\partial R^2|_{R=R_m}>0$,
we find the relative equilibrium distance between two black holes corresponding to the minimum of $V$:
\begin{eqnarray}
R_m=\frac{L^2}{G\mu^2 M}.
\label{eq_pot4n}
\end{eqnarray}
%
We assume that the total angular momentum  $J=GM^2/c$
of di-black-hole system \cite{Murad1,Murad2,Murad3}
is conserved during the conservative mass transfer and
the orbital angular momentum of the  di-black-hole is
\begin{eqnarray}
L=J-S_{1}-S_{2}=\frac{G(M^2-M_{1}^2-M_{2}^2)}{c}.
\label{eq_pot4nL}
\end{eqnarray}
Here the orbital and spin angular momenta are the parallel and the orbital angular momentum is minimal.
Employing Eqs. (\ref{eq_pot4n}) and (\ref{eq_pot4nL}), we derive the expression  for   relative distance
\begin{eqnarray}
R_m=\frac{4GM}{c^2}.
\label{eq_pot4n2}
\end{eqnarray}
 As seen, the larger $M$ the larger $R_m$ is. Because $R_m/(R_1+R_2)=4$,
the separation of components increases at  $M_1\to M_2$.
From Eqs.  (\ref{eq_pot3}), (\ref{eq_pot3n}), and (\ref{eq_pot4n2}) one can obtain the simple formula
\begin{eqnarray}
V(R_m)=-\frac{GM_1M_2}{2R_m}=-\frac{M_1M_2c^2}{8M}=-\frac{\mu c^2}{8}
\label{eq_pot5nn}
\end{eqnarray}
for the interaction potential.
Here, $V(R_m)$ depends only on the reduced mass  $\mu$ and  velocity of light.


So, using Eqs. (\ref{eq_pot1}), (\ref{eq_pot1R}), and (\ref{eq_pot5nn}),  we obtain the final expression
\begin{eqnarray}
U=-\frac{Mc^2}{2}\left(1+\frac{M_1M_2}{4M^2}\right)=-\frac{Mc^2}{2}\left(1+\frac{\mu}{4M}\right)
\label{eq_pot2}
\end{eqnarray}
for the total potential energy of the binary black hole system.
For the BBH considered, $v(R_m)=(GM/R_m)^{1/2}=c/2$  and  with a satisfactory accuracy one can
neglect the relativistic effects and  use the Newtonian law of gravity.

Using the mass asymmetry coordinate $\eta$ instead of masses $M_1=\frac{M}{2}(1+\eta)$ and $M_2=\frac{M}{2}(1-\eta)$,
we rewrite   the   expressions (\ref{eq_pot4nL}) and  (\ref{eq_pot2}) for the orbital angular momentum
\begin{eqnarray}
L=\frac{GM^2}{2c}\left(1-\eta^2\right)
\label{eq_pot7nn}
\end{eqnarray}
and the total potential energy
\begin{eqnarray}
U=-\frac{Mc^2}{2}\left(1+\frac{1-\eta^2}{16}\right).
\label{eq_pot7n}
\end{eqnarray}

%
%

Since the solution of equation
$$\frac{\partial U}{\partial\eta}=\frac{Mc^2}{16}\eta=0$$
gives $\eta=\eta_m=0$
and
$$\frac{\partial^2 U}{\partial\eta^2}|_{\eta=\eta_m}=\frac{Mc^2}{16} > 0,$$
the potential landscape has an oscillator shape and
a global minimum at $\eta=\eta_m=0$   for the arbitrary total mass $M$ of the BBH system (Fig. 1). 
So,   the transfer of mass between the black holes in the binary system
is energetically favorable and, correspondingly, can occur. The initial asymmetric system
is driven without additional driving energy to the symmetric BBH configuration
(towards a global minimum).
As seen, this conclusion does not depend on the choice of parameters, and the losses of the total mass and orbital angular momentum
do not influence the symmetrization (the transfer of mass from the heavier component to the light one) of system.
Due to the interaction energy between two black holes with their own extreme gravity fields,
it becomes possible to transfer matter between them.
%
%
However,    the evolution of binary system
in  mass asymmetry coordinate depends also on the mass parameter in this degree of freedom.
Since the surfaces of two black holes are spaced $R_1 + R_2 < R_m$ from each other, the mass parameter in
$\eta$ is very large and, accordingly, to some extent prohibits the symmetrization of the di-black-hole system.
\begin{figure}[h]
\includegraphics[scale=0.7]{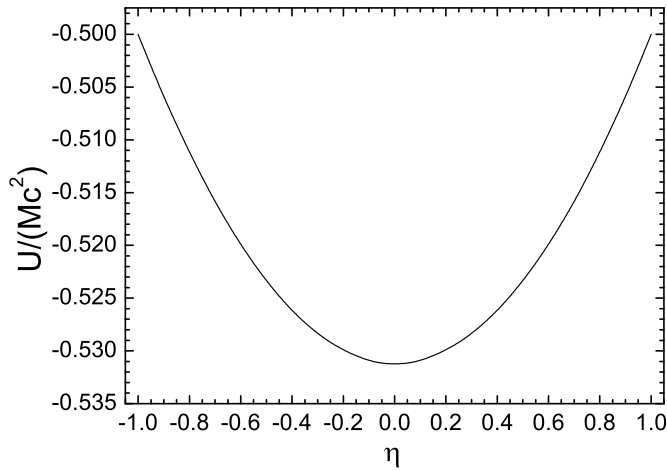}
\vspace*{0.5cm}
\caption{The potential energy of the BBH (14) divided by $Mc^2$ as a function of mass asymmetry $\eta$.
}
\label{n_time}
\end{figure}
\begin{figure}[h]
\includegraphics[scale=0.7]{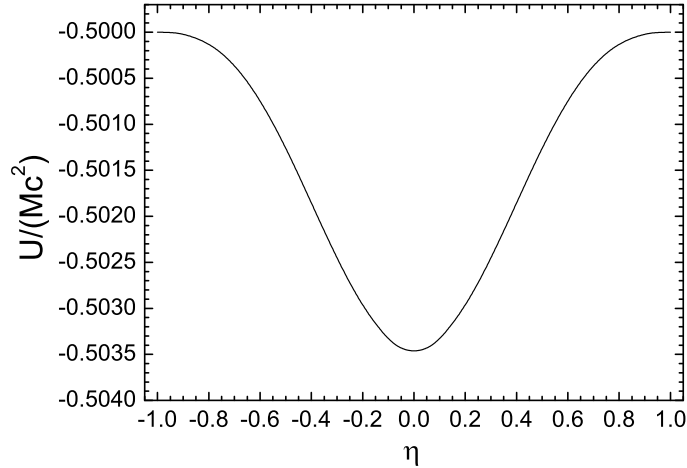}
\vspace*{0.5cm}
\caption{The potential energy of the BBH (16) divided by $Mc^2$ as a function of mass asymmetry $\eta$.
}
\label{n_time}
\end{figure}


The  asymmetrization (merger due to the transfer of mass from the lighter component to the heavy one)
 of the binary black holes considered
is not energetically favorable process and, correspondingly, the merger channel in mass asymmetry is strongly suppressed for di-black-hole.
Thus,  the question remains open on the mechanism
of the merger of   two black holes and the origin of gravitational waves   \cite{Abbott}.

In the case of the antiparallel orbital and spin angular momenta, the orbital angular momentum
\begin{eqnarray}
L=J+S_{1}+S_{2}=\frac{G(M^2+M_{1}^2+M_{2}^2)}{c}
\label{eq_pot4nL2}
\end{eqnarray}
is the maximal.
For the binary black hole system with the orbital angular momentum (\ref{eq_pot4nL2}),
the total potential energy is
\begin{eqnarray}
U=-\frac{Mc^2}{2}\left(1+\frac{(1-\eta^2)^3}{16(3+\eta^2)^2}\right)
\label{eq_pot9n}
\end{eqnarray}
and  $R_m/(R_1+R_2)\ge 36$, $v(R_m)=(GM/R_m)^{1/2}\le c/6$, $\eta_m=0$, $\partial^2 U/\partial\eta^2|_{\eta=\eta_m}> 0$,
  and all conclusions given above are also valid in this case.
  As seen from the comparison of Fig. 2 with Fig. 1, 
  the same conclusions are also obtained
  in the case when the black holes of binary system are spinning in opposite directions.

\section{Summary}\label{sec:Summary}
Employing the Newtonian law of gravity and considering the potential energy of the
di-black-hole as a function of mass asymmetry (transfer) collective coordinate $\eta$,
we have shown the possibility of matter transfer between the black holes in a binary system.
%
%
The   evolution
of asymmetric binary system to the symmetric configuration with $\eta=\eta_m=0$ is
energetically favorable process.
Although black holes have their own strong gravitational fields, the transfer of matter between
two black holes arises from the energy of interaction between them.
A di-black-hole system does not send any signals  during its evolution. One can indirectly observe
only the result of this evolution. In the course of time evolution the maximum of the distribution of expected BBH mass ratios
shifts towards binary systems with equal masses.

The symmetrization
of   initially asymmetric  BBH leads to the decrease of potential energy $U$,
thus transforming   the potential energy into internal kinetic energy.
For example, for the binary black-holes $4M_{\odot}+2M_{\odot}$ ($\eta_i=0.33$)  and
$36M_{\odot}+29M_{\odot}$ ($\eta_i=0.11$),
the internal energies of  black-holes
will
increase during symmetrization
by the amount
$\Delta U=U(\eta_i)-U(\eta=0)=Mc^2\eta_i^2\approx 10^{47}$ J. For the comparison,
 in the cases of compact binary  stars and compact binary galaxies, the released energies are about  $10^{41}$~J  and $10^{48-52}$~J,
 respectively \cite{IJMPE,IJMPE2}. So, the BBH is the source of thermal energy.


The transfer of matter from a lighter component to a heavy one, leading to the merger of black holes in a binary system,
is not an energetically favorable process and, accordingly, the question of the origin of gravitational waves remains open.



In the frame of our model, one can also perform the dynamic calculations of the evolution of binary system
in  mass asymmetry coordinate.
But this extension of our model is the subject  of future studies.

\section*{Acknowledgements}

This work was partially supported by  Russian Foundation for Basic Research (Moscow)  and
DFG (Bonn). V.V.S.   acknowledge  the partial
 supports from the Alexander von Humboldt-Stiftung (Bonn).


\end{document}